\documentclass[twocolumn,english,aps,pra]{revtex4-1}

\usepackage{amsmath,amsthm,amssymb}
\usepackage{graphicx,epstopdf}
\usepackage{color,bm,babel}
\usepackage[unicode=true,pdfusetitle,
 bookmarks=true,bookmarksnumbered=false,bookmarksopen=false,
 citecolor=cyan,urlcolor=magenta,linkcolor=blue, breaklinks=false,
 pdfborder={0 0 0},backref=false,colorlinks=true]{hyperref}



\begin{document}

\title{Orbital Angular Momentum Induced by Nonabsorbing
Optical Elements through Space-variant Polarization-state Manipulations}

\author{Dengke Zhang }
\email{dkzhang@outlook.com}

\author{Xue Feng}
\email{x-feng@tsinghua.edu.cn }

\author{Yidong Huang}

\affiliation{Tsinghua National Laboratory for Information Science and Technology,
Department of Electronic Engineering, Tsinghua University, Beijing
100084, China}
\begin{abstract}
To manipulate orbital angular momentum (OAM) carried by light beams,
there is a great interest in designing various optical elements from
the deep-ultraviolet to the microwave. Normally, the OAM variation
introduced by optical elements can be attributed to two terms, namely
the dynamic and geometric phases. Up till now, the dynamic contribution
induced by optical elements has been clearly recognized. However,
the contribution of geometric phase still seems obscure, especially
considering the vector vortex beams. In this work, an analytical formula
is derived to fully describe the OAM variation introduced by the nonabsorbing
optical elements, which perform space-variant polarization-state manipulations.
It is found that the geometric contribution can be further divided
into two parts: one is directly related to optical elements and the
other one explicitly relies solely on the vortices before and after
the transformations. Based on this result, the same OAM variation
can be achieved with different combinations of the dynamic and/or
geometric contributions. With numerical simulations, it is shown that
transformation of the optical vortices can be fully and flexibly designed
with a family of optical elements. We believe that these results are
helpful to understand the effect of optical elements and offer a new
perspective to design the optical elements for manipulating the OAM
carried by light beams. 
\end{abstract}

\keywords{orbital angular momentum; optical vortices; optical elements; geometric
phase; Pancharatnam phase; Poincar\'{e} sphere; Stokes parameters}
\maketitle

\section{Introduction}

Light can carry both spin and orbital angular momentum (SAM and OAM),
which are corresponding to the polarization and spatial degrees of
freedom, respectively \cite{Padgett2004PT,Franke-Arnold2008LPR,Barnett2002JOB}.
Under the paraxial approximation, the SAM and OAM are separable within
isotropic homogeneous media \cite{Yao2011AOP}. The SAM per photon
has a value of $\pm\hbar$ (the reduced Planck's constant) corresponding
to left-/right-handed circular polarization, while the OAM would be
more intriguing even under paraxial approximation. For a scalar vortex
beam, the OAM would be $l\hbar$ per photon for the optical field
with a spiral wavefront of $\exp(il\phi)$, where $l$ can be any
integer \cite{Allen1992PRA}. However, for vector vortex beams, it
would be more complicated since the space-variant state of polarization
(SOP) would attribute to the OAM charge \cite{Wang2010PRL,Niv2006OE}.
To address it, several approaches have been proposed to extract the
geometric contribution through the high-order Poincar\'{e} spheres
\cite{Milione2011PRL,Milione2012PRL} or introducing the topological
Pancharatnam charge \cite{Zhang2015SR}. However, when the light beam
is transformed, there are no explicit formulas to describe the corresponding
variation of OAM charge due to the geometric contribution. Such an
explicit formula would be significant while analyzing the OAM evolution
in an optical system and tailoring the OAM carried by vortex beams,
since the spin-orbit interactions (SOIs) are inevitable. 

The SOI is a general basic phenomenon in manipulations of light beams
and photons, which has been observed in light propagating \cite{Bliokh2008NP,Bliokh2017PRL},
scattering \cite{OConnor2014NC}, focusing \cite{Zhao2007PRL}, etc.
The SOIs have evoked some interesting investigations of physical phenomena
such as the spin-Hall effect \cite{Onoda2004PRL,Hosten2008S,Bliokh2015S},
extraordinary momentum states \cite{Bliokh2014NCa} and even extended
to cavity-quantum electrodynamics (CQEDs) \cite{Deng2014PRL}. In
the new reality of nano-optics, SOI is essential in both the physical
conception and device design and should be also taken into account
for nano-optical systems. Recently, several nano-optics platforms
have been employed to replicate the functionality of common optical
elements such as polarizers, wave retarders, etc., which have shown
promising abilities to manipulate both polarization and phase distributions
of optical beams \cite{Arbabi2015NN}. In particular, SOI has emerged
as a powerful mean to tailor the OAM carried by scalar vortex beams,
which can be achieved by optical elements to perform space-variant
polarization-state manipulations (e.g. spiral phase plates \cite{Yu2011S},
q-plates \cite{Marrucci2006PRL,Bomzon2002OLa}, J-plates \cite{Devlin2017S}).
In these transformations, the desired spiral wavefronts of light beams
are introduced by steering dynamic phase and/or geometric phase. However,
the SOIs would be much more complicated while considering vector vortex
beams passing through optical elements, where the geometric phase
has to be seriously considered to evaluate the OAM of light beams
\cite{Souza2007PRL,Ma2016NC}. Furthermore, more interesting phenomena
and flexible manipulations of optical vortices can be achieved with
SOIs in inhomogeneous or anisotropic media. The manipulation of SOIs
can release the full potential of information processing through an
effective utilization of both SAM and OAM. Thus the generation, measurement,
and control of optical vortices via SOIs have attracted a considerable
amount of attentions recently. Definitely, two cruxes, namely OAM
variation and geometric phase, are inevitable in the SOIs of optical
vortices. Thus, there are two questions that should be addressed.
First, whether the OAM variation introduced by optical element is
distinguishable in terms of dynamic and geometric phases for arbitrary
vortex beams? And second, whether there are various designs of optical
elements to tailor OAM as steering SAM? These issues are quite appealing
for both the theoretical understanding and practical application. 

In this work, we have tackled both issues. An explicit formula is
deduced to describe the OAM variation introduced by the nonabsorbing
optical elements, which perform space-variant polarization-state manipulations.
It is found that the geometric contribution can be further divided
into two parts: one is directly related to optical elements and the
other one explicitly relies solely on the optical vortices before
and after the transformations. Specifically, an intuitive picture
is presented to obtain deeper insight into how the dynamic and geometric
phases are involved. As a concrete example, we present the design
rule for transformations from one scalar vortex to another to show
the flexibility for the same OAM variation. Furthermore, the designs
of transforming vector vortex are also shown. At the end, the features
of previously reported optical elements would be discussed under our
theoretical framework. 

\section{Theoretical principle}

\subsection{OAM of an optical vortex beam}

Under the paraxial approximation, an electric field of a fully polarized
vector vortex beam propagating along $\mathbf{z}$ direction with
the angular frequency $\omega$ can be written as \cite{Torres2011Book}

\begin{equation}
\mathbf{E}\left(x,y\right)=i\omega\left(\alpha\mathbf{\hat{x}}+\beta\mathbf{\hat{y}}+\frac{i}{k}\left(\frac{\partial\alpha}{\partial x}+\frac{\partial\beta}{\partial y}\right)\mathbf{\hat{z}}\right)e^{ikz},\label{EqEf}
\end{equation}

\noindent where $\alpha(\beta)$ represents the complex amplitude
of $x(y)-$component of electric field, as a function of $(x,y)$
(omitted for simplicity). Then the SOP of such a beam can be described
by a $2\times1$ Jones vector $\left|a\right\rangle =(a_{x},a_{y})^{\mathrm{T}}$,
where $a_{x}(a_{y})=\alpha(\beta)/\sqrt{I_{\text{E}}}$ represents
the normalized complex amplitude and $I_{\text{E}}=|\alpha|^{2}+|\beta|^{2}$
is the electric intensity of light. The corresponding Stokes vector
$\mathbf{S=}$$(S_{1},S_{2},S_{3})^{\mathrm{T}}$ is defined by $\ensuremath{S_{j}=\left\langle a|\boldsymbol{\sigma}_{j}|a\right\rangle (j=1,2,3)}$,
where $\boldsymbol{\sigma}_{j}$ are the Pauli matrices and $S_{0}=\left\langle a|\boldsymbol{\sigma}_{0}|a\right\rangle $,
where $\boldsymbol{\sigma}_{0}$ equals $2\times2$ identity matrix
\cite{Gutierrez-Vega2011OL}. Thus, $S_{0}=1$ presents fully polarized
light and $S_{3}=\pm1$ presents left/right circularly polarized field
$\left|\mathrm{e}_{\pm}\right\rangle =(1/\sqrt{2})(1,\pm i)^{\mathrm{T}}$.
By mapping $\mathbf{S}$ directly in three-dimensional Cartesian coordinates,
the Poincar\'{e} sphere can be constructed and the corresponding
azimuth ($\psi_{\mathrm{S}}$) and ellipticity ($\chi_{\mathrm{S}}$)
angles of SOP can be resolved by $\tan(2\psi_{\mathrm{S}})=S_{2}/S_{1}$
and $\sin(2\chi_{\mathrm{S}})=S_{3}/S_{0}$ \cite{Born2013Book}. 

With the aforementioned notations, the average OAM charge for a fully
polarized paraxial vector vortex beam can be calculated by OAM density
$\mathbf{j}_{z}^{\mathrm{o}}$ as \cite{Allen2000OC,Barnett2002JOB}

\begin{align}
\bar{l} & =\frac{\iint\mathbf{j}_{z}^{\text{o}}rdrd\phi}{\omega\epsilon_{0}\iint I_{\text{E}}S_{0}rdrd\phi}.\label{EqLavg}
\end{align}

\noindent It should be noticed that the average OAM charge depends
on not only the distribution of SOP but also the intensity distribution
($I_{\text{E}}$) of beams. Thus, equation \eqref{EqLavg} is also
applicable to characterize the non-vortex (‘asymmetry’) OAM of beams
without the wavefront singularities \cite{Bekshaev2003JOSAA}. Next,
similar to our previous work \cite{Zhang2015SR}, the OAM density
can be expressed by introducing the Pancharatnam connection between
two different SOPs \cite{Berry1987JMO}. Here, circularly polarized
fields are adopted as reference. Then, the phase difference for any
field $\left|a\right\rangle =(a_{x},a_{y})^{\mathrm{T}}$ can be written
as $\psi_{\mathrm{P\pm}}=\arg\left(\left\langle \mathrm{e}_{\pm}|a\right\rangle \right).$
According to Ref.~\cite{Zhang2015SR}, the OAM density can be obtained
by

\begin{equation}
\frac{\mathbf{j}_{z}^{\text{o}}}{\omega\epsilon_{0}I_{\text{E}}}=\frac{1}{2}\left((1+S_{3})\frac{\partial\psi_{\mathrm{P+}}}{\partial\phi}+(1-S_{3})\frac{\partial\psi_{\mathrm{P-}}}{\partial\phi}\right).\label{EqLdens}
\end{equation}
The detailed deduction of equations \eqref{EqLavg} and \eqref{EqLdens}
can be found in Appendix A. 

In equation \eqref{EqLdens}, the derivative of $\psi_{\mathrm{P\pm}}$
is known as the topological Pancharatnam charge \cite{Niv2006OE,Zhang2015SR}.
With equations \eqref{EqLavg} and \eqref{EqLdens}, the average OAM
charge carried by the light beam can be fully expressed with the SAM
($S_{3}$) and the topological Pancharatnam charge ($\partial\psi_{\mathrm{P\pm}}/\partial\phi$),
which can depict the OAM states on a single Poincar\'{e} sphere as
Refs.~\cite{Zhang2015SR,Zhao2017SR}. Thus, the corresponding geometric
phase for any transformations can be conveniently identified on the
same Poincar\'{e} spheres. Such a representation can succinctly and
elegantly describe the OAM of a light beam, where the contribution
from the space-variant SOP of vector vortex has been naturally embedded.
Moreover, the OAM charge can be identified with standard measurement
of Stokes parameters and interferometry. As shown in the following
part, our approach could be conveniently employed to design optical
elements for manipulating the OAM charge and investigate the OAM evolution
of light beam propagating in an optical system. 

\begin{figure}
\includegraphics{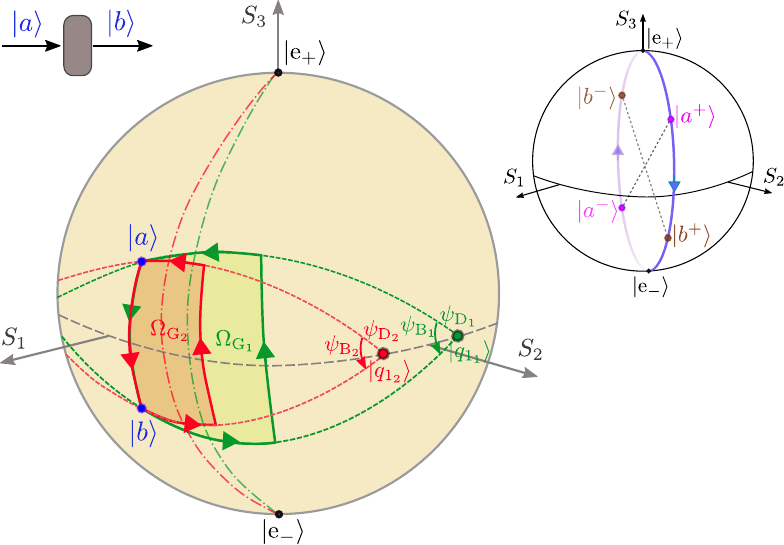}\caption{Manipulating both SAM and OAM. For the same transformation from $|a\rangle$
to $|b\rangle$, different linear operations (i.e., with different
Jones matrices $\mathbf{J}_{i}$, where $i=1,2$) can induce distinct
OAM charge since there are different combinations of dynamic phases
$\psi_{\mathrm{D}_{i}}$ and geometric phases $\Omega_{\mathrm{G}_{i}}/4$
($i=1,2$). The magnitude of $\Omega_{\mathrm{G}_{i}}$ depends on
the eigen-polarization $|q_{1_{i}}\rangle$ and birefringent phase
difference $\psi_{\mathrm{B}_{i}}$ of $\mathbf{J}_{i}$. The inset
shows that one linear operation can transform a pair of orthogonal
SOP scalar vortices $|a^{\pm}\rangle$ into another pair of SOP scalar
vortices $|b^{\pm}\rangle$ with different OAM variations at the same
time. \label{Fig1.P2PScheme} }
\end{figure}

\subsection{OAM variation induced by non-absorbing optical elements}

\begin{figure}
\includegraphics{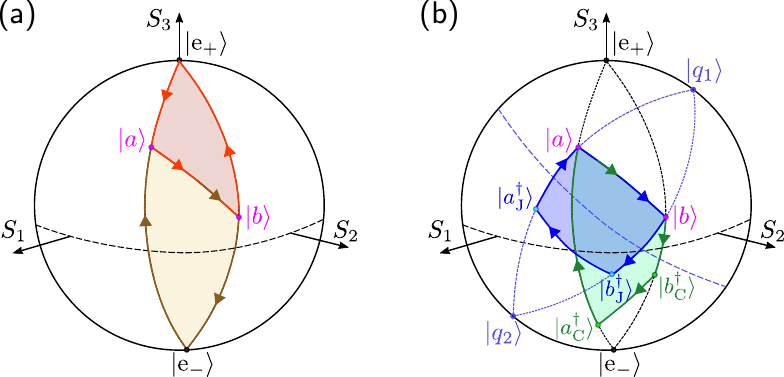}\caption{Solid angle associated with geometric phase in the transformation.
(a) Geodesic triangle $\Omega_{ab\mathrm{e}_{+}}^{\mathrm{C}}$ (reddish)
and $\Omega_{ab\mathrm{e}_{-}}^{\mathrm{C}}$ (buff) on the Poincar\'{e}
sphere. (b) Spherical quadrangle $\Omega_{abb^{\dagger}a^{\dagger}}^{\mathrm{J}}$
(blue) on the Poincar\'{e} sphere, which is a portion of lune of
dihedral angle defined by states $|a\rangle$, $|b\rangle$ and $|q_{1(2)}\rangle$.
The state $|a_{\mathrm{J}}^{\dagger}\rangle(|b_{\mathrm{J}}^{\dagger}\rangle)$
is a reflection of state $|a\rangle(|b\rangle)$ referring to mirror
plane of a great circle, which is perpendicular to the axis joining
the states $|q_{1}\rangle$ and $|q_{2}\rangle$. Similarly, spherical
quadrangle $\Omega_{abb^{\dagger}a^{\dagger}}^{\mathrm{C}}$ (green)
is a portion of lune of dihedral angle defined by states $|a\rangle$,
$|b\rangle$ and $|\mathrm{e}_{\pm}\rangle$. The state $|a_{\mathrm{C}}^{\dagger}\rangle(|b_{\mathrm{C}}^{\dagger}\rangle)$
is a reflection of state $|a\rangle(|b\rangle)$ referring to mirror
plane of the equator. \label{Fig2.Solidangle}}
\end{figure}

Here, we consider a scenario that a light passes through a nonabsorbing
optical element. The SOPs of input and output fields are denoted as
$\left|a\right\rangle =(a_{x},a_{y})^{\mathrm{T}}$ and $\left|b\right\rangle =(b_{x},b_{y})^{\mathrm{T}}$,
respectively. And the optical element is characterized by a unitary
Jones matrix $\mathbf{J}$, i.e., $\mathbf{J}^{\dagger}=\mathbf{J}^{-1}$.
Thus, the light field transformation can be described as $|b\rangle=\mathbf{J}|a\rangle$
(see Fig.~\ref{Fig1.P2PScheme}). Mathematically, the eigenvalues
and eigenstates of $\mathbf{J}$ are $\mu_{1(2)}$ and $|q_{1(2)}\rangle$,
respectively. Then the corresponding Stokes vectors for eigenstates
can be calculated as $\mathbf{S}^{\mathrm{J}}=\mathbf{S}^{q_{1}}=-\mathbf{S}^{q_{2}}=(S_{1}^{\mathrm{J}},S_{2}^{\mathrm{J}},S_{3}^{\mathrm{J}})^{\mathrm{T}}$,
where $S_{j}^{\mathrm{J}}=\left\langle q_{1}|\boldsymbol{\sigma}_{j}|q_{1}\right\rangle $.
With these notations, the variation of OAM density can be deduced
according to equation \eqref{EqLdens}. For transforming state $\left|a\right\rangle $
to state $\left|b\right\rangle $, beyond SAM variation from $S_{3}^{a}$
to $S_{3}^{b}$, there is also a variation from $\psi_{\mathrm{P\pm}}^{a}$
to $\psi_{\mathrm{P\pm}}^{b}$, where the superscript $a(b)$ refers
to the parameters related to state $|a\rangle$($|b\rangle$). According
to Refs.~\cite{Gutierrez-Vega2011OL,Martinez-Fuentes2012OC}, the
phase difference $\psi_{\mathrm{P\pm}}^{a\rightarrow b}=\psi_{\mathrm{P\pm}}^{b}-\psi_{\mathrm{P\pm}}^{a}$
can be rewritten as

\begin{equation}
\psi_{\mathrm{P\pm}}^{a\rightarrow b}=\psi_{\mathrm{D}}-\frac{\Omega_{ab\mathrm{e}_{\pm}}^{\mathrm{C}}}{2}+\frac{\Omega_{abb^{\dagger}a^{\dagger}}^{\mathrm{J}}}{4},\label{EqdPanp}
\end{equation}

\noindent where $\psi_{\mathrm{D}}=\mathrm{arg}(\mu_{1}\mu_{2})/2$
presents the dynamic phase as the light beam propagating through the
optical element and $\Omega_{ab\mathrm{e}_{\pm}}^{\mathrm{C}}\big/2$
is the geometric phase introduced by varied SOP between the output
and input fields, which corresponds to parallel transport of the state
around a closed loop ($|a\rangle\rightarrow|b\rangle\rightarrow|\mathrm{e}_{\pm}\rangle\rightarrow|a\rangle$)
on the Poincar\'{e} sphere (see Fig.\textbf{~}\ref{Fig2.Solidangle}(a)).
While $\Omega_{abb^{\dagger}a^{\dagger}}^{\mathrm{J}}$ is a spherical
quadrangle corresponding to the closed trajectory $|a\rangle\rightarrow|b\rangle\rightarrow|b_{\mathrm{J}}^{\dagger}\rangle\rightarrow|a_{\mathrm{J}}^{\dagger}\rangle\rightarrow|a\rangle$,
as blue area shown in Fig.~\ref{Fig2.Solidangle}(b) (also see Fig.~\ref{fig:S01JM}),
where $|a_{\mathrm{J}}^{\dagger}\rangle(|b_{\mathrm{J}}^{\dagger}\rangle)$
holds the Stokes vector $\mathbf{S}^{a_{\mathrm{J}}^{\dagger}(b_{\mathrm{J}}^{\dagger})}=\mathbf{S}^{a(b)}-2(\mathbf{S}^{a(b)}\cdot\mathbf{S}^{\mathrm{J}})\mathbf{S}^{\mathrm{J}}$.
It can be found that the term $\Omega_{abb^{\dagger}a^{\dagger}}^{\mathrm{J}}\big/4$
is the geometric phase explicitly related with the optical element
$\mathbf{J}$. It should be noticed that although both $\Omega_{ab\mathrm{e}_{\pm}}^{\mathrm{C}}\big/2$
and $\Omega_{abb^{\dagger}a^{\dagger}}^{\mathrm{J}}\big/4$ are related
to the geometric phases, they would affect the final OAM density with
different manners. To clearly describe the contribution of optical
element and the impact of varied SOP between input and output fields,
the variation of OAM density can be deduced with equation \eqref{EqdPanp}
as follows:

\begin{align}
\frac{\Delta\mathbf{j}_{z}^{\text{o}}}{\omega\epsilon_{0}I_{\text{E}}}= & \frac{\partial\psi_{\mathrm{D}}}{\partial\phi}+\frac{\partial}{\partial\phi}\left(\frac{\Omega_{abb^{\dagger}a^{\dagger}}^{\mathrm{J}}}{4}\right)\nonumber \\
 & +\left[S_{3}^{a}\frac{\partial\psi_{\mathrm{S}}^{a}}{\partial\phi}-S_{3}^{b}\frac{\partial\psi_{\mathrm{S}}^{b}}{\partial\phi}-\frac{\partial}{\partial\phi}\left(\frac{\Omega_{abb^{\dagger}a^{\dagger}}^{\mathrm{C}}}{4}\right)\right],\label{EqdLdens}
\end{align}

\noindent where $\Omega_{abb^{\dagger}a^{\dagger}}^{\mathrm{C}}$
is a spherical quadrangle defined by states $|a\rangle$, $|b\rangle$,
$|b_{\mathrm{C}}^{\dagger}\rangle$ and $|a_{\mathrm{C}}^{\dagger}\rangle$
as green area shown in Fig.\textbf{~}\ref{Fig2.Solidangle}(b), where
$|a_{\mathrm{C}}^{\dagger}\rangle(|b_{\mathrm{C}}^{\dagger}\rangle)$
holds the Stokes vector $\mathbf{S}^{a_{\mathrm{C}}^{\dagger}(b_{\mathrm{C}}^{\dagger})}=\mathbf{S}^{a(b)}-2(\mathbf{S}^{a(b)}\cdot\mathbf{S}^{\mathrm{e}_{+}})\mathbf{S}^{\mathrm{e}_{+}}$.
It is easy to find $\Omega_{abb^{\dagger}a^{\dagger}}^{\mathrm{C}}=\Omega_{ab\mathrm{e}_{+}}^{\mathrm{C}}+\Omega_{ab\mathrm{e}_{-}}^{\mathrm{C}}$
(see Appendix B for details). According to equation \eqref{EqdLdens},
the OAM variation can be attributed to three terms. The first term
($C_{\mathrm{d}}=\frac{\partial\psi_{\mathrm{D}}}{\partial\phi}$)
is dynamic contribution and presents the OAM variation induced by
the dynamic phase delay, which only depends on $\psi_{\mathrm{D}}$
of the optical element ($\mathbf{J})$, regardless of the SOP of input
beam. The rest two terms present the geometric contributions ($C_{\mathrm{g}}$)
that rely on the optical elements as well as the SOP of light beams.
Specifically, the second term ($C_{\mathrm{g}}^{\mathrm{J}}=\frac{1}{4}\frac{\partial}{\partial\phi}\left(\Omega_{abb^{\dagger}a^{\dagger}}^{\mathrm{J}}\right)$)
is related to eigen-polarization $\mathbf{\mathbf{S}^{\mathrm{J}}}$
(i.e. $|q_{1}\rangle$) and birefringent phase difference $\psi_{\mathrm{B}}$
(equals $\mathrm{arg}(\mu_{1}^{\ast}\mu_{2})$) of the adopted transformation
matrix $\mathbf{J}$ (see corresponding spherical quadrangle $\Omega_{\mathrm{G}_{i}}$
in Fig.\textbf{~}\ref{Fig1.P2PScheme} or $\Omega_{abb^{\dagger}a^{\dagger}}^{\mathrm{J}}$
in Fig.\textbf{~}\ref{Fig2.Solidangle}(b)). The third (rest) term
($C_{\mathrm{g}}^{\mathrm{V}}$) explicitly depends on the input and
output fields themselves and presents the geometric contribution stemming
from the different SOP distributions of input and output vortices.
Namely $C_{\mathrm{g}}^{\mathrm{V}}$ can be fully determined by $\mathbf{S}^{a}$
and $\mathbf{S}^{b}$ (for $\Omega_{abb^{\dagger}a^{\dagger}}^{\mathrm{C}}$,
see Fig.\textbf{~}\ref{Fig2.Solidangle}(b)). Thus, the geometric
contribution of $C_{\mathrm{g}}^{\mathrm{V}}$ would be determined
once the input and target output vortex beams are given. However,
there are still various combinations of dynamic ($C_{\mathrm{d}}$)
and geometric ($C_{\mathrm{g}}^{\mathrm{J}}$) contributions to achieve
the same OAM variation. Thus, equation \eqref{EqdLdens} indicates
that it would be greatly flexible to design the optical element for
vortex beam transformations. It should be mentioned that this has
not been fully perceived and explored at present. To demonstrate the
mentioned above, some simulations have been carried out for both the
scalar and vector vortex beams.

\section{Transformations on scalar vortices}

\begin{figure*}
\includegraphics{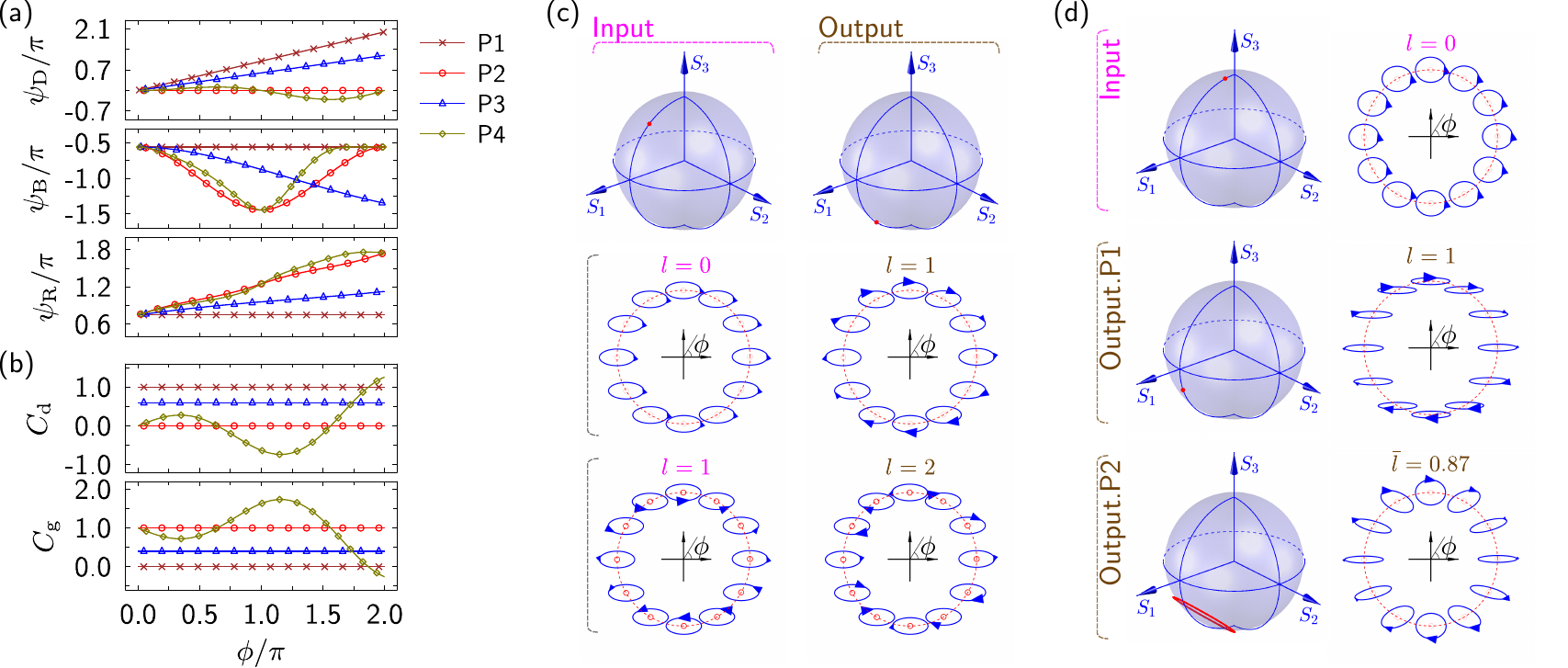}\caption{Transformations on scalar vortices. (a) The design parameters $\{\psi_{\mathrm{D}},\psi_{\mathrm{B}},\psi_{\mathrm{R}}\}$
of Jones matrices for optical plates P1--P4 from the top down. (b)
The contributions for OAM variations from dynamic term $C_{\mathrm{d}}$
and geometric term $C_{\mathrm{g}}$ for designed input SOP $\{2\psi_{\mathrm{S}},2\chi_{\mathrm{S}}\}=\{0,50^{\circ}\}$.
Note that we set $I_{\mathrm{E}}(\phi)=1$. (c) SOP trace on the Poincar\'{e}
sphere (top panel) and spatial distribution of SOP (middle and bottom
panels) for input (left panel) and output (right panel) fields with
a designed P2P transformation. For the same input SOP but different
OAM charges ($l=0$ (middle panel) or $l=1$ (bottom panel)), they
can be transferred to both the same SOP and OAM variation by any of
P1--P4. (d) SOP trace on the Poincar\'{e} sphere (left panel) and
spatial distribution of SOP (left panel) for input fields with SOP
$\{2\psi_{\mathrm{S}},2\chi_{\mathrm{S}}\}=\{0,80^{\circ}\}$ (top
panel). The transferred fields are not the same due to different attributes
of dynamic (P1, middle panel) and geometric (P2, bottom panel) contributions.\label{Fig3.P2PTrans} }
\end{figure*}

First, the point-to-point (P2P) transformation is demonstrated on
the Poincar\'{e} sphere for a scalar vortex beam with the same OAM
variation but different designs, as sketched in Fig.~\ref{Fig1.P2PScheme}.
According to equation \eqref{EqdLdens}, it can be found that $C_{\mathrm{g}}^{\mathrm{V}}=0$
for P2P transformation of scalar vortex. Thus there are two contributions
for the OAM variation. The first one is the dynamic contribution ($C_{\mathrm{d}}$),
which is determined by $\psi_{\mathrm{D}}$ of optical elements. The
second term is geometric contribution ($C_{\mathrm{g}}^{\mathrm{J}}$),
which stems from geometric phase $\Omega_{\mathrm{G}}/4$ depending
on the $|q_{1}\rangle$ and $\psi_{\mathrm{B}}$. As a scalar vortex,
the input light beam can be fully described by SOP of $\{2\psi_{\mathrm{S}},2\chi_{\mathrm{S}}\}$
and OAM charge of $l$. It should be noted that the SOPs of the scalar
vortices are space-invariant, so $\{2\psi_{\mathrm{S}},2\chi_{\mathrm{S}}\}$
are $(x,y)$-independent. For the sake of simplicity but without loss
of generality, $\psi_{\mathrm{S}}=0$ is settled since the absolute
azimuth angle is irrelevant due to the rotation symmetry of the coordinate.
Thus, the input scalar vortex can be expressed as $|a^{+}\rangle=e^{il\phi}(\cos(\chi_{\mathrm{S}}),i\sin(\chi_{\mathrm{S}}))^{\mathrm{T}}$.
The output vortex ($|b^{+}\rangle$) is considered as $|b^{+}\rangle=e^{i(l+\Delta l)\phi}(\cos(\chi_{\mathrm{S}}),-i\sin(\chi_{\mathrm{S}}))^{\mathrm{T}}$
with a flipped handedness and OAM variation of $\Delta l$ after a
optical element, which is described by Jones matrix $\mathbf{J}$
(see the inset of Fig.~\ref{Fig1.P2PScheme}). It is easy to find
that, for such a transformation $|b^{+}\rangle=\mathbf{J}|a^{+}\rangle$,
the term $e^{il\phi}$ can be canceled so that the transformation
is independent on OAM charge of input beam. Generally, $\mathbf{J}$
would be linearly birefringent without considering chiral or magneto-optic
materials, thus the eigenstates of $\mathbf{J}$ are two linear and
orthogonal eigen-polarizations. Considering the unitary nature of
$\mathbf{J}$, the eigenvalues of $\mathbf{J}$ are given by $e^{i\psi_{\mathrm{D}}}\{e^{-i\psi_{\mathrm{B}}/2},e^{i\psi_{\mathrm{B}}/2}\}$
with dynamic phase delay $\psi_{\mathrm{D}}$ and birefringent phase
difference $\psi_{\mathrm{B}}$. And the orthogonal eigen-polarizations
can be written as $\mathbf{R}(\psi_{\mathrm{R}})(1,0)^{\mathrm{\mathrm{T}}}$
and $\mathbf{R}(\psi_{\mathrm{R}})(0,1)^{\mathrm{T}}$, where $\mathbf{R(\cdot)}$
is the standard rotation matrix and $\psi_{\mathrm{R}}$ is the orientation
angle of linear eigen-polarizations. Thus, for the considered optical
elements, the transformation matrix $\mathbf{J}$ can be determined
by three parameters $\{\psi_{\mathrm{D}}(\phi),\psi_{\mathrm{B}}(\phi),\psi_{\mathrm{R}}(\phi)\}$
with each $\phi$ 

\begin{widetext}

\begin{equation}
\mathbf{J}=e^{i\psi_{\mathrm{D}}}\left(\begin{array}{cc}
\cos\left(\frac{\psi_{\mathrm{B}}}{2}\right)-i\sin\left(\frac{\psi_{\mathrm{B}}}{2}\right)\cos\left(2\psi_{\mathrm{R}}\right) & -i\sin\left(\frac{\psi_{\mathrm{B}}}{2}\right)\sin\left(2\psi_{\mathrm{R}}\right)\\
-i\sin\left(\frac{\psi_{\mathrm{B}}}{2}\right)\sin\left(2\psi_{\mathrm{R}}\right) & \cos\left(\frac{\psi_{\mathrm{B}}}{2}\right)+i\sin\left(\frac{\psi_{\mathrm{B}}}{2}\right)\cos\left(2\psi_{\mathrm{R}}\right)
\end{array}\right).
\end{equation}

\end{widetext}

It should be noticed that there are only two equations ($|b^{+}\rangle=\mathbf{J}|a^{+}\rangle$)
to confine the relations of such three parameters. Thus, there would
be various strategies to set the $\mathbf{J}$ with the same transformation
result. In other words, once the dynamic phase delay $\psi_{\mathrm{D}}(\phi)$
is assigned, a combination of $\{\psi_{\mathrm{B}}(\phi),\psi_{\mathrm{R}}(\phi)\}$
can always be found. Obviously, it is a family of optical plates to
perform the P2P transformation that only relies on input SOP, regardless
of the carried OAM charge (details are discussed in Appendix C). To
demonstrate such unique feature, some simulations have been carried
out. 

In Fig.~\ref{Fig3.P2PTrans}, four different optical plates (denoted
as P1--P4) are designed to transform left-handed elliptical polarization
($\{2\psi_{\mathrm{S}},2\chi_{\mathrm{S}}\}=\{0,50^{\circ}\}$) vortex
to right-handed elliptical polarization ($\{2\psi_{\mathrm{S}},2\chi_{\mathrm{S}}\}=\{0,-50^{\circ}\}$)
vortex with $\Delta l=1$. Figure \ref{Fig3.P2PTrans}(a) shows the
parameters of $\mathbf{J}$ for each optical plate and Fig.~\ref{Fig3.P2PTrans}(b)
shows the corresponding dynamic term $C_{\mathrm{d}}$ and geometric
term $C_{\mathrm{g}}^{\mathrm{J}}$ that would induce the OAM variations
(it should be noted that $C_{\mathrm{g}}^{\mathrm{V}}=0$). For P1
and P2, the OAM variation is purely induced by dynamic or geometric
contributions, respectively. Both $\psi_{\mathrm{B}}$ and $\psi_{\mathrm{R}}$
keep constant for P1 while $\psi_{\mathrm{D}}$ keeps constant for
P2. As a comparison, both dynamic and geometric terms would contribute
to the OAM variation for P3 and P4. Both $C_{\mathrm{d}}$ and $C_{\mathrm{g}}$
are designed as homogeneous and inhomogeneous distribution along $\phi$
for P3 and P4, respectively. Obviously, P4 is a more general and flexible
example. Moreover, the corresponding SOP on the Poincar\'{e} sphere
and the electric field distribution have been calculated for each
optical plate to verify the equivalence of the considered transformations.
As expected, the final results are the same for P1--P4 as shown in
Fig.~\ref{Fig3.P2PTrans}(c). It can also be found that the same
OAM variation ($\Delta l=1$) can be obtained with input of $l=0$
or $l=1$ for P1--P4. It coincides with that the OAM variation is
independent of the input OAM charge. 

It should be mentioned that if the SOP of input light changes, P1--P4
will introduce different transformations since the geometric contribution
depends on the SOP of the input beam. Thus, different designs of optical
plates would introduce diverse variations of SAM and OAM when the
SOP of input beam does not match the designed one. Figure \ref{Fig3.P2PTrans}(d)
shows the transformations for input plane-wave with polarization $\{2\psi_{\mathrm{S}},2\chi_{\mathrm{S}}\}=\{0,80^{\circ}\}$
by P1 and P2. For P1, the output is still a scalar vortex with $l=1$
since there is the only pure dynamic contribution as shown in Fig.~\ref{Fig3.P2PTrans}(d).
However, the average OAM variation $\Delta\bar{l}$ would equal 0.87
for P2 and the output would be a vector vortex as shown in Fig.~\ref{Fig3.P2PTrans}(d).
The reason is that both two geometrical terms would contribute to
the OAM variation (see Fig.~\ref{fig:S02}). For the orthogonal input
SOPs (antipodal points on the Poincar\'{e} sphere), there are equal
but opposite geometric contributions since they have opposite evolution
direction on the Poincar\'{e} sphere.\emph{ }Thus, the same dynamic
term $C_{\mathrm{d}}$ and opposite geometric term $C_{\mathrm{g}}$
would be introduced and the final result is $\Delta\bar{l}\propto C_{\mathrm{d}}\pm C_{\mathrm{g}}$.
For a desired OAM variation with the given SOP, the introduced contributions
can be dynamic and/or geometric. Thus, the design of P2P transformations
is flexible and fully controllable according to requirements. But
it should be noticed that the dynamic phase based optical elements
have a SOP-independent response while geometric phase based optical
elements are completely SOP-dependent. Thus, the SOP-bandwidth of
the optical elements would be narrower if more geometric contribution
is introduced. Such issue should be considered for specific applications.

\section{Transformations on vector vortices}

\begin{figure}
\includegraphics{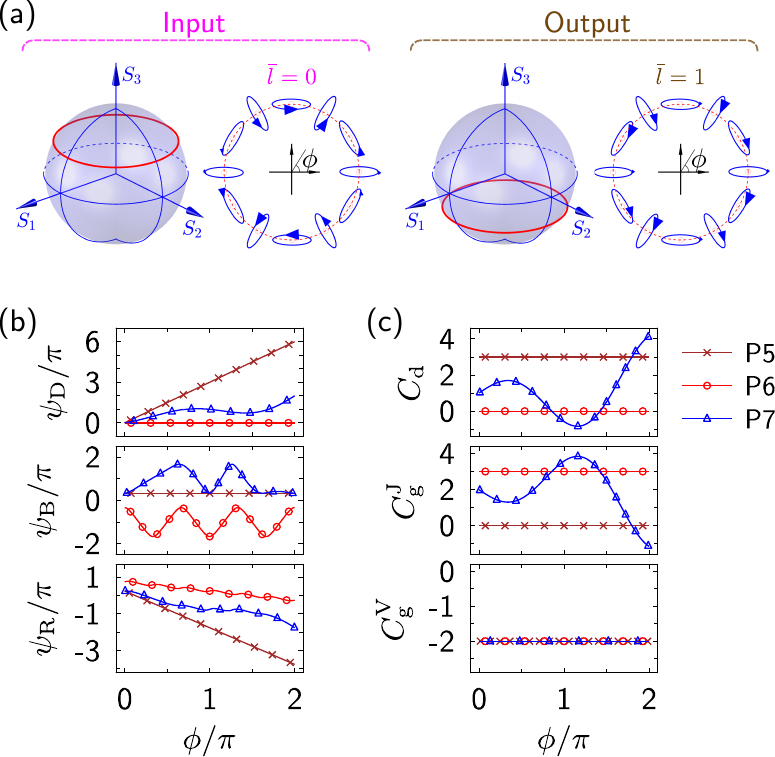}

\caption{Transformations on vector vortex. (a) SOP trace on the Poincar\'{e}
sphere and spatial distribution of SOP for input (left panel) and
output (right panel) vector vortices with a designed transformation
by any of optical plates P5--P7. (b) The design parameters $\{\psi_{\mathrm{D}},\psi_{\mathrm{B}},\psi_{\mathrm{R}}\}$
of Jones matrices for optical plates P5--P7 from the top down. (c)
The contributions for OAM variation from dynamic term $C_{\mathrm{d}}$,
geometric term $C_{\mathrm{g}}^{\mathrm{J}}$ and $C_{\mathrm{g}}^{\mathrm{V}}$
for input vector vortex shown in (a) from the top down. The contribution
from $C_{\mathrm{g}}^{\mathrm{V}}$ are the same for any of P5--P7.
Note that we set $I_{\mathrm{E}}(\phi)=1$. \label{Fig4.VVTrans}}
\end{figure}

More generally, equation \eqref{EqdLdens} can be applied on transforming
vector vortices. As shown in Fig.~\ref{Fig4.VVTrans}(a), the input
beam is a cylindrical vortex with $2\chi_{\mathrm{S}}=30^{\circ}$
and $\bar{l}=0$ while the output beam is another cylindrical vortex
with $2\chi_{\mathrm{S}}=-30^{\circ}$ and $\bar{l}=1$ (see Appendix
D and Fig.~\ref{fig:S03} for $\{2\psi_{\mathrm{S}},2\chi_{\mathrm{S}}\}$
of the input and output beams). For such a transformation, the linearly
birefringent unitary $\mathbf{J}$ is employed. By numerically solving
the $\mathbf{J}$, the design parameters for three different optical
plates (named P5--P7) are displayed in Fig.~\ref{Fig4.VVTrans}(b).
The corresponding dynamic and geometric contributions are given in
Fig.~\ref{Fig4.VVTrans}(c). It can be found that the values of $C_{\mathrm{g}}^{\mathrm{V}}$
are the same but not equal to zero. For these cases, the optical elements
have to be meticulously designed to achieve average OAM variation
of $\Delta\bar{l}=1$. Similar to P2P transformations, P5 is designed
with only dynamic contribution $C_{\mathrm{d}}$ and P6 is with only
geometric contribution $C_{\mathrm{g}}^{\mathrm{J}}$. Moreover, both
two terms are designed for P7. Though the designs are not so straightforward
as that for P2P transformation, the portions of dynamic and geometric
contributions are quantitatively controllable by careful design of
optical elements with equation \eqref{EqdLdens}, which is very important
to the modern precise measurement and control. It should be noticed
that there is no theoretical limitation for applying equation \eqref{EqdLdens}
on designing optical elements. However, in reality, it is not easy
to achieve arbitrary transformation on vector vortices since the physically
implemented Jones matrices would be limited by the available materials
and structures.

\section{Discussion}

\begin{figure}
\includegraphics{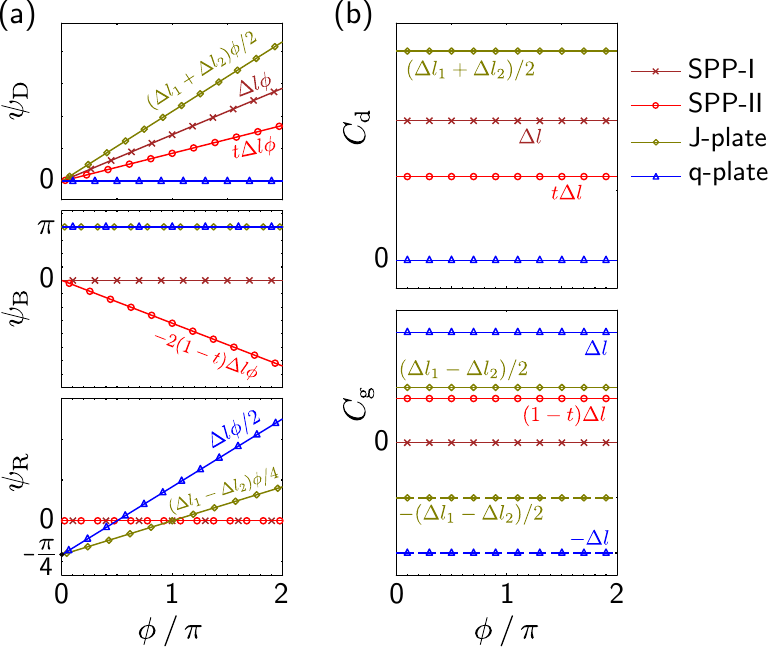}

\caption{Features of spiral phase plates (SPPs), J-plates, and q-plates. (a)
Design parameters $\{\psi_{\mathrm{D}},\psi_{\mathrm{B}},\psi_{\mathrm{R}}\}$
of Jones matrices from the top down. For SPP-I and SPP-II, the desired
OAM variation is $\Delta l$ and the portion from dynamic contribution
is set as $t$ for SPP-II. For J-plate, the desired OAM variation
is $\Delta l_{1/2}$ for input light with left-/right-handed circular
polarization. For q-plate, the desired OAM variation is $\pm\Delta l$
for input light with left-/right-handed circular polarization. (b)
The corresponding dynamic (upper panel) and geometric (lower panel)
contributions for OAM variation. For J-plate and q-plate, the opposite
geometric contributions will be induced according to handedness of
input light. \label{Fig5.JQSplate}}
\end{figure}

So far, there are three common optical plates---spiral phase plates,
q-plates, and J-plates, which have been employed to generate and manipulate
OAM beams \cite{Oemrawsingh2004AO,Yu2011S,Devlin2017S,Biener2002OL,Marrucci2006PRL,Karimi2014LSA}.
Actually, for all of them, the operation mechanism can be understood
and explained by our theoretical approach. Here, some discussions
and comments will be given. For spiral phase plates (SPPs), there
are two types according to equation \eqref{EqdLdens}. The first one
(SPP-I) is fabricated with homogeneous materials and the required
phase delay is introduced by the spiral design of the plate \cite{Oemrawsingh2004AO}.
Thus, there is only the pure dynamic contribution so that the same
dynamic phase, as well as the same OAM variation, can be obtained
in spite of the SOP of input beams. For the second type (SPP-II),
the desired OAM variation can be achieved only for a specific SOP
of the input beam while the output beam would have the same SOP \cite{Yu2011S}.
Thus, SPP-II is a particular case of P2P transformation, where the
SOP of the input beam is just coincided with one of the eigen-polarizations
of the plate and the SOP would be maintained. Since it just works
for a specific SOP, it is not a pure dynamic phase based optical plate.
Actually, for SPP-II, both the dynamic and geometric contributions
have to be taken into account for the OAM variations case by case.
Obviously, the operation mechanism of SPP-II is totally different
with SPP-I since the OAM variation is not solely introduced by dynamic
contribution. 

For the J-plate \cite{Devlin2017S}, it could be considered as a special
type of P2P transformation. The special constraint is that the orthogonal
SOPs of input beam should be transferred to a flipped handedness with
a different OAM variation at the same time. As shown in the inset
of Fig.~\ref{Fig1.P2PScheme}, for a J-plate, the OAM variation $\Delta l_{1}$
and $\Delta l_{2}$ should be obtained for $|b^{+}\rangle=\mathbf{J}|a^{+}\rangle$
and $|b^{-}\rangle=\mathbf{J}|a^{-}\rangle$, respectively and simultaneously.
According to our framework, it means that the dynamic contributions
are always the same but the geometric contributions are opposite for
orthogonal input SOPs. Thus, the J-plate can be designed as that the
dynamic OAM variation is $(\Delta l_{1}+\Delta l_{2})/2$ and the
opposite geometric OAM variation should be $\pm(\Delta l_{1}-\Delta l_{2})/2$
according to the handedness of SOP of the input beam. Then the combinations
of three parameters $\{\psi_{\mathrm{D}},\psi_{\mathrm{B}},\psi_{\mathrm{R}}\}$
can be readily obtained. Particularly, if the input field is circular
polarization ($2\chi_{\mathrm{S}}=\pm\pi/2$), three parameters would
hold simple relations as shown in Fig.~\ref{Fig5.JQSplate}(a). This
kind of J-plates is a half-plate and can flip circular SOP with different
OAM variations. Specifically, if $\Delta l_{1}=-\Delta l_{2}=\Delta l$,
it is the well-known q-plate, in which only the pure geometric contribution
is introduced \cite{Biener2002OL,Marrucci2006PRL,Karimi2014LSA} (see
Appendix C for details). The full parameters of $\mathbf{J}$ for
these mentioned optical plates and the corresponding contributions
of each term for OAM variations are summarized and presented in Fig.~\ref{Fig5.JQSplate}. 

As a summary, this work presents an explicit formula to evaluate the
OAM variation due to the optical elements in terms of both dynamic
and geometric phases. With the help of the topological Pancharatnam
charge, the geometric phases can be further separated into two parts.
One is directly related to optical elements and the other one solely
relies on SOPs of the input and output light beams. Such treatment
is not just a mathematical trick but would introduce a new viewpoint
to fully understand the operation mechanism and would be helpful to
explore the flexibility of designing the optical elements according
to the applications. For instance, pure dynamic contribution based
optical plates can implement identical OAM variations in spite of
the SOP of input beam while pure geometric contribution based optical
plates can serve as a mode sorter for both SAM and OAM in modern optical
communication systems. Moreover, our theoretical approach can be employed
for the optical systems to analyze influences due to the dynamic and
geometric phases. In this work, only the case of linear orthogonal
eigen-polarizations of the Jones matrix is considered since it is
the common response of most materials and structures. It should be
mentioned that our theoretical approach is not limited by this constraint.
Actually, if the eigen-polarizations of the Jones matrix could be
arbitrary, more complicated functions can be achieved for various
potential applications. Additionally, there are several assumptions
in our theoretical deduction such as unitary Jones matrix, paraxial
beam, and fully polarized fields. Actually, breaking either of them
would introduce some more interesting investigations, e.g. considering
inhomogeneous Jones matrix \cite{Cerjan2017PRL}, non-Hermitian (including
\emph{PT}-symmetry) systems \cite{Lawrence2014PRL}, or non-reciprocal
systems \cite{Mahmoud2015NC}. Furthermore, only classical light fields
are considered in our work, but we believe that the similar work about
quantum counterpart would bring more things of new physics and our
work could evoke some fundamental research about spin-orbit interaction
and the related topics. 

\section*{Acknowledgment}

This work was supported by the National Key Research and Development
Program of China (Grant No. 2017YFA0303700), the National Natural
Science Foundation of China (Grant No. 61621064), Beijing Innovation
Center for Future Chip, and Beijing Academy of Quantum Information
Science.

\bigskip{}

\renewcommand{\theequation}{A\arabic{equation}} 
\setcounter{equation}{0} 

\section*{Appendix A: Orbital angular momentum of an optical vortex}

Under the paraxial approximation, the electric and magnetic fields
of a fully polarized vector vortex beam of angular frequency $\omega$
propagating along $\mathbf{z}$ direction can be written as \cite{Torres2011Book}

\begin{subequations} 

\begin{align*}
\mathbf{E}\left(x,y\right) & =i\omega\left(\alpha\mathbf{\hat{x}}+\beta\mathbf{\hat{y}}+\frac{i}{k}\left(\frac{\partial\alpha}{\partial x}+\frac{\partial\beta}{\partial y}\right)\mathbf{\hat{z}}\right)e^{ikz},\\
\mathbf{B}\left(x,y\right) & =ik\left(-\beta\mathbf{\hat{x}}+\alpha\mathbf{\hat{y}}+\frac{i}{k}\left(-\frac{\partial\beta}{\partial x}+\frac{\partial\alpha}{\partial y}\right)\mathbf{\hat{z}}\right)e^{ikz},
\end{align*}

\end{subequations}

\noindent where $\alpha$ and $\beta$ represent the complex amplitude
of $x-$ and $y-$component of electric field, respectively. They
can be written as

\begin{subequations}

\begin{align*}
\alpha(x,y) & =\sqrt{I_{\text{E}}(x,y)}a_{x}(x,y),\\
\beta(x,y) & =\sqrt{I_{\text{E}}(x,y)}a_{y}(x,y),
\end{align*}

\end{subequations}

\noindent where $a_{x}(a_{y})=\alpha(\beta)/\sqrt{I_{\text{E}}}$
are normalized electric field components with electric intensity of
$I_{\text{E}}=|\alpha|^{2}+|\beta|^{2}$. So the polarization state
of this light at each site can be described by a $2\times1$ Jones
vector $\left|a\right\rangle =(a_{x},a_{y})^{\mathrm{T}}$. Then,
Stokes vector $\mathbf{S}=(S_{1},S_{2},S_{3})^{\mathrm{T}}$ is defined
by $S_{j}=\left\langle a|\boldsymbol{\sigma}_{j}|a\right\rangle (j=1,2,3)$,
where $\boldsymbol{\sigma}_{j}$ are the Pauli matrices \cite{Gutierrez-Vega2011OL},

\[
\boldsymbol{\sigma}_{1}=\begin{pmatrix}1 & 0\\
0 & -1
\end{pmatrix},\bm{\quad}\boldsymbol{\sigma}_{2}=\begin{pmatrix}0 & 1\\
1 & 0
\end{pmatrix},\bm{\quad}\boldsymbol{\sigma}_{3}=\begin{pmatrix}0 & -i\\
i & 0
\end{pmatrix}.
\]

Meanwhile, $S_{0}=\left\langle a|\boldsymbol{\sigma}_{0}|a\right\rangle $,
where $\boldsymbol{\sigma}_{0}$ equals $2\times2$ identity matrix.
Thus, for left/right circularly polarized light field $\left|\mathrm{e}_{\pm}\right\rangle =(1/\sqrt{2})(1,\pm i)^{\mathrm{T}}$,
there is $S_{3}=\pm1$. Then plotting Stokes vector $\mathbf{S}$
on three-dimensional Cartesian coordinates, the Poincar\'{e} sphere
could be constructed and the corresponding azimuth ($\psi_{\mathrm{S}}$)
and ellipticity ($\chi_{\mathrm{S}}$) angles are resolved by, respectively

\begin{subequations} 

\begin{align}
\tan(2\psi_{\mathrm{S}}) & =S_{2}/S_{1},\label{S4a}\\
\sin(2\chi_{\mathrm{S}}) & =S_{3}/S_{0}.\label{S4b}
\end{align}

\end{subequations} 

\begin{widetext}

The linear momentum density, which is defined as $\mathbf{p}=\varepsilon_{0}\mathbf{E}\times\mathbf{B}$,
can be expressed and divided into transverse and longitudinal components,

\begin{subequations} 

\begin{align}
\mathbf{p}_{\bot} & =i\frac{\omega\varepsilon_{0}}{2}\left[\left(\alpha\nabla\alpha^{*}+\beta\nabla\beta^{*}-\alpha^{*}\nabla\alpha-\beta^{*}\nabla\beta\right)+2\nabla\times\left(\left(\alpha^{*}\beta-\beta^{*}\alpha\right)\hat{\mathbf{z}}\right)\right],\label{S5a}\\
\mathbf{p}_{z} & =\omega k\varepsilon_{0}\left(\left|\alpha\right|^{2}+\left|\beta\right|^{2}\right)=\omega k\varepsilon_{0}I_{\text{E}}S_{0}.\label{S5b}
\end{align}

\end{subequations}

Meanwhile, the energy density of such a beam is

\begin{equation}
w=c\mathbf{p}_{z}=\varepsilon_{0}\omega^{2}\left(\left|\alpha\right|^{2}+\left|\beta\right|^{2}\right)=\varepsilon_{0}\omega^{2}I_{\text{E}}S_{0}.\label{S6}
\end{equation}

Then, the cross product of linear momentum density with $\mathbf{r}$
(radius vector) gives the angular momentum density, so $z-$component
of angular momentum density is 

\begin{equation}
\mathbf{j}_{z}=\left(\mathbf{r}\times\mathbf{p}\right)_{z}=i\frac{\omega\varepsilon_{0}}{2}\left[\left(\alpha\frac{\partial}{\partial\phi}\alpha^{*}+\beta\frac{\partial}{\partial\phi}\beta^{*}-\alpha^{*}\frac{\partial}{\partial\phi}\alpha-\beta^{*}\frac{\partial}{\partial\phi}\beta\right)+2r\frac{\partial}{\partial r}\left(\alpha^{*}\beta-\beta^{*}\alpha\right)\right].\label{S7}
\end{equation}

Further, $\mathbf{j}_{z}$ can be divided into spin and orbital parts
as 

\begin{subequations} 

\begin{align}
\mathbf{j}_{z}^{\mathrm{s}} & =i\omega\varepsilon_{0}r\frac{\partial}{\partial r}\left(\alpha^{*}\beta-\beta^{*}\alpha\right)=\omega\varepsilon_{0}r\frac{\partial\left(I_{\mathrm{E}}S_{3}\right)}{\partial r},\label{S8a}\\
\mathbf{j}_{z}^{\mathrm{o}} & =i\frac{\omega\varepsilon_{0}}{2}\left(\alpha\frac{\partial}{\partial\phi}\alpha^{*}+\beta\frac{\partial}{\partial\phi}\beta^{*}-\alpha^{*}\frac{\partial}{\partial\phi}\alpha-\beta^{*}\frac{\partial}{\partial\phi}\beta\right).\label{S8b}
\end{align}

\end{subequations} 

\end{widetext}

With the ratio of angular momentum to energy that is examined by Allen
\cite{Allen2000OC}, the average SAM charge and OAM charge can be
calculated as

\begin{subequations}\label{S9} 

\begin{align}
\bar{s} & =\omega\frac{\iint\mathbf{j}_{z}^{\text{s}}rdrd\phi}{\iint wrdrd\phi}=\frac{\iint I_{\text{E}}S_{3}rdrd\phi}{\iint I_{\text{E}}S_{0}rdrd\phi},\label{S9a}\\
\bar{l} & =\omega\frac{\iint\mathbf{j}_{z}^{\text{o}}rdrd\phi}{\iint wrdrd\phi}=\frac{\iint\mathbf{j}_{z}^{\text{o}}rdrd\phi}{\omega\epsilon_{0}\iint I_{\text{E}}S_{0}rdrd\phi}.\label{S9b}
\end{align}

\end{subequations} 

Then, we introduce the phase difference for two different SOPs of
$\left|\mathrm{e}_{\mathrm{A}}\right\rangle $ and $\left|\mathrm{e}_{\mathrm{B}}\right\rangle $
through the Pancharatnam connection, which is defined by \cite{Berry1987JMO}

\begin{equation}
\psi_{\mathrm{P}}=\arg\left(\left\langle \mathrm{e}_{\mathrm{A}}|\mathrm{e}_{\mathrm{B}}\right\rangle \right).\label{S10}
\end{equation}

Here, using left/right circularly polarized fields as reference fields,
the phase difference for any field $\left|a\right\rangle =(a_{x},a_{y})^{\mathrm{T}}$
can be written as

\begin{equation}
\psi_{\mathrm{P\pm}}=\arg\left(\left\langle \mathrm{e}_{\pm}|a\right\rangle \right).\label{S11}
\end{equation}

According to Ref.~\cite{Zhang2015SR}, we can obtain

\begin{equation}
\frac{\mathbf{j}_{z}^{\text{o}}}{\omega\epsilon_{0}I_{\text{E}}}=S_{0}\frac{\partial\psi_{\mathrm{P\pm}}}{\partial\phi}\pm(S_{0}\mp S_{3})\frac{\partial\psi_{\mathrm{S}}}{\partial\phi},\label{S16}
\end{equation}
then using equation \eqref{S16}, we can get a relation

\begin{equation}
\frac{\partial\psi_{\mathrm{S}}}{\partial\phi}=-\frac{1}{2}\left(\frac{\partial\psi_{\mathrm{P+}}}{\partial\phi}-\frac{\partial\psi_{\mathrm{P-}}}{\partial\phi}\right),\label{S17}
\end{equation}
then equation \eqref{S16} can be rewritten as

\begin{equation}
\frac{\mathbf{j}_{z}^{\text{o}}}{\omega\epsilon_{0}I_{\text{E}}}=\frac{1}{2}\left((S_{0}+S_{3})\frac{\partial\psi_{\mathrm{P+}}}{\partial\phi}+(S_{0}-S_{3})\frac{\partial\psi_{\mathrm{P-}}}{\partial\phi}\right).\label{S18}
\end{equation}

Substituting equation \eqref{S16} or \eqref{S18} into equation \eqref{S9b},
we can calculate the average OAM charge for any vortex beams. In equations
\eqref{S16} and \eqref{S18}, the derivative of $\psi_{\mathrm{P\pm}}$
is known as the topological Pancharatnam charge. With equation \eqref{S16},
we have found that the OAM of a vector vortex can be divided into
two parts: the topological Pancharatnam charge and contribution from
geometric phase induced by space-variant SOP of light fields, which
is consistent with the reported results \cite{Wang2010PRL,Niv2006OE}
and more detailed discussions were provided in our previous work \cite{Zhang2015SR}. 

\section*{Appendix B: OAM variation induced by optical elements}

Here, we consider a scenario that the light beam passes through a
nonabsorbing optical element and investigate the OAM variation induced
by this optical element, which is characterized by a unitary Jones
matrix $\mathbf{J}$ (i.e., $\mathbf{J}^{\dagger}=\mathbf{J}^{-1}$)
with the eigenvalues of $\mu_{1(2)}$, eigenstates of $|q_{1(2)}\rangle$,
and the corresponding Stokes vectors of $\mathbf{S}^{\mathrm{J}}=\mathbf{S}^{q_{1}}=-\mathbf{S}^{q_{2}}=(S_{1}^{\mathrm{J}},S_{2}^{\mathrm{J}},S_{3}^{\mathrm{J}})^{\mathrm{T}}$
($S_{j}^{\mathrm{J}}=\left\langle q_{1}|\boldsymbol{\sigma}_{j}|q_{1}\right\rangle $)
\cite{Gutierrez-Vega2011OL}. When light field $|a\rangle$ passes
through the optical element $\mathbf{J}$, the output beam can be
expressed as $|b\rangle=\mathbf{J}|a\rangle$. From equation \eqref{S9b},
it could be known that the variation of OAM simply depends on the
variation of $\mathbf{j}_{z}^{\text{o}}$ due to non-absorbing nature
($I_{\mathrm{E}}^{a}=I_{\mathrm{E}}^{b}=I_{\mathrm{E}}$). Then with
equations \eqref{S17} and \eqref{S18}, the variation of $\mathbf{j}_{z}^{\text{o}}$
can be deduced as

\begin{widetext}

\begin{equation}
\frac{\Delta\mathbf{j}_{z}^{\text{o}}}{\omega\epsilon_{0}I_{\text{E}}}=\frac{1}{2}\left((S_{3}^{b}-S_{3}^{a})\frac{\partial\psi_{\mathrm{P+}}^{a}}{\partial\phi}+(S_{0}^{b}+S_{3}^{b})\frac{\partial\psi_{\mathrm{P+}}^{a\rightarrow b}}{\partial\phi}+(S_{3}^{a}-S_{3}^{b})\frac{\partial\psi_{\mathrm{P-}}^{a}}{\partial\phi}+(S_{0}^{b}-S_{3}^{b})\frac{\partial\psi_{\mathrm{P-}}^{a\rightarrow b}}{\partial\phi}\right),\label{S19}
\end{equation}

\end{widetext}

\noindent where the superscript $a(b)$ refers to the parameters related
to state $|a\rangle$($|b\rangle$) and $\psi_{\mathrm{P\pm}}^{a\rightarrow b}=\psi_{\mathrm{P\pm}}^{b}-\psi_{\mathrm{P\pm}}^{a}$
is the difference of phases (defined by the Pancharatnam connection)
between $|a\rangle$ and $|b\rangle$. According to Refs.~\cite{Gutierrez-Vega2011OL,Martinez-Fuentes2012OC},
$\psi_{\mathrm{P\pm}}^{a\rightarrow b}$ can be written as

\begin{equation}
\psi_{\mathrm{P\pm}}^{a\rightarrow b}=\psi_{\mathrm{D}}-\frac{\Omega_{ab\mathrm{e}_{\pm}}^{\mathrm{C}}}{2}+\frac{\Omega_{abb^{\dagger}a^{\dagger}}^{\mathrm{J}}}{4},\label{S20}
\end{equation}
where $\psi_{\mathrm{D}}=\mathrm{arg}(\mu_{1}\mu_{2})/2$ is dynamic
phase gained by the beam when it propagates through the optical element,
$\Omega_{ab\mathrm{e}_{\pm}}^{\mathrm{C}}\big/2$ is the geometric
phase, related to the referenced circularly polarized field, which
corresponds to parallel transport of the state around a closed loop
($|a\rangle\rightarrow|b\rangle\rightarrow|\mathrm{e}_{\pm}\rangle\rightarrow|a\rangle$)
on the Poincar\'{e} sphere (see Fig.\textbf{~}\ref{Fig2.Solidangle}(a)
in main text), and $\Omega_{abb^{\dagger}a^{\dagger}}^{\mathrm{J}}\big/4$
is the geometric phase introduced by the optical element $\mathrm{\mathbf{J}}$.
For the third term, $\Omega_{abb^{\dagger}a^{\dagger}}^{\mathrm{J}}$
is a spherical quadrangle corresponding to the closed trajectory $|a\rangle\rightarrow|b\rangle\rightarrow|b_{\mathrm{J}}^{\dagger}\rangle\rightarrow|a_{\mathrm{J}}^{\dagger}\rangle\rightarrow|a\rangle$,
as shown in Fig.~\ref{Fig2.Solidangle}(b) in main text, where $|a_{\mathrm{J}}^{\dagger}\rangle(|b_{\mathrm{J}}^{\dagger}\rangle)$
holds the Stokes vector $\mathbf{S}^{a_{\mathrm{J}}^{\dagger}(b_{\mathrm{J}}^{\dagger})}=\mathbf{S}^{a(b)}-2(\mathbf{S}^{a(b)}\cdot\mathbf{S}^{\mathrm{J}})\mathbf{S}^{\mathrm{J}}$.

Further, using equations \eqref{S17} and \eqref{S20}, the equation
\eqref{S19} can be rewritten as

\begin{widetext}

\begin{equation}
\frac{\Delta\mathbf{j}_{z}^{\text{o}}}{\omega\epsilon_{0}I_{\text{E}}}=S_{0}^{b}\frac{\partial\psi_{\mathrm{D}}}{\partial\phi}+S_{0}^{b}\frac{\partial}{\partial\phi}\left(\frac{\Omega_{abb^{\dagger}a^{\dagger}}^{\mathrm{J}}}{4}\right)+\left[S_{3}^{a}\frac{\partial\psi_{\mathrm{S}}^{a}}{\partial\phi}-S_{3}^{b}\frac{\partial\psi_{\mathrm{S}}^{b}}{\partial\phi}-S_{0}^{b}\frac{\partial}{\partial\phi}\left(\frac{\Omega_{abb^{\dagger}a^{\dagger}}^{\mathrm{C}}}{4}\right)\right],\label{S21}
\end{equation}

\end{widetext}

\noindent where $\Omega_{abb^{\dagger}a^{\dagger}}^{\mathrm{C}}$
is a spherical quadrangle defined by states $|a\rangle$, $|b\rangle$,
$|b_{\mathrm{C}}^{\dagger}\rangle$ and $|a_{\mathrm{C}}^{\dagger}\rangle$
as shown in Fig.\textbf{~}\ref{Fig2.Solidangle}(b) in main text,
where $|a_{\mathrm{C}}^{\dagger}\rangle(|b_{\mathrm{C}}^{\dagger}\rangle)$
holds the Stokes vector $\mathbf{S}^{a_{\mathrm{C}}^{\dagger}(b_{\mathrm{C}}^{\dagger})}=\mathbf{S}^{a(b)}-2(\mathbf{S}^{a(b)}\cdot\mathbf{S}^{\mathrm{e}_{+}})\mathbf{S}^{\mathrm{e}_{+}}$.
Thus using equation \eqref{S9b}, the variation of OAM charge can
be solved

\begin{equation}
\Delta\bar{l}=\frac{\iint I_{\mathrm{E}}\Delta\mathbf{j}_{z}^{\text{o}}rdrd\phi}{\omega\epsilon_{0}\iint I_{\mathrm{E}}S_{0}^{a}rdrd\phi}.\label{S21b}
\end{equation}

\noindent It should be noticed that equation \eqref{S21b} is applicable
to the transformation performed by nonabsorbing optical elements.

Note that, for any spherical triangle defined by states $|a\rangle$,
$|b\rangle$ and $|c\rangle$, whose Stokes vectors are $\mathbf{S}^{a}$,
$\mathbf{S}^{b}$ and $\mathbf{S}^{c}$, respectively, the triangular
area $\Omega_{abc}$ is 

\noindent 
\begin{equation}
\Omega_{abc}=2\mathrm{arctan}\left[\frac{\mathbf{S}^{a}\cdot(\mathbf{S}^{b}\times\mathbf{S}^{c})}{1+\mathbf{S}^{a}\cdot\mathbf{S}^{b}+\mathbf{S}^{b}\cdot\mathbf{S}^{c}+\mathbf{S}^{c}\cdot\mathbf{S}^{a}}\right].\label{A01}
\end{equation}

\begin{figure}
\includegraphics{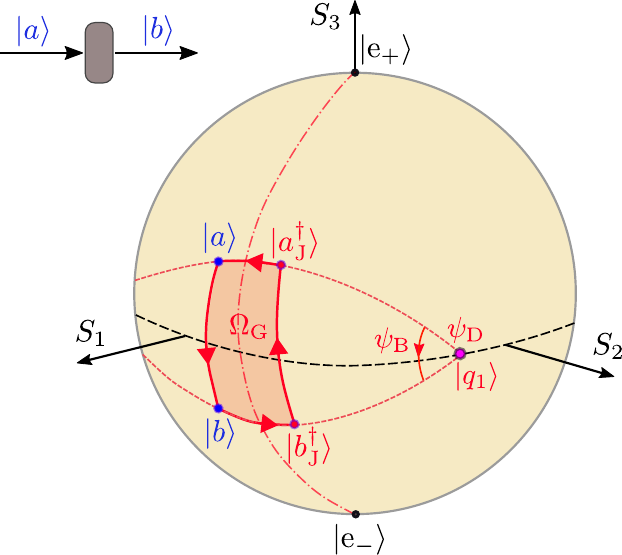}\caption{For the transformation from $|a\rangle$ to $|b\rangle$, there will
be the introduced dynamic phase $\psi_{\mathrm{D}}$ and geometric
phase $\Omega_{\mathrm{G}}/4$ by an optical element of $\{\psi_{\mathrm{D}},\psi_{\mathrm{B}}\}$
and eigen-polarizations $|q_{1(2)}\rangle$. $\Omega_{\mathrm{G}}=\Omega_{abb^{\dagger}a^{\dagger}}^{\mathrm{J}}$
is a spherical quadrangle corresponding to the closed trajectory $|a\rangle\rightarrow|b\rangle\rightarrow|b_{\mathrm{J}}^{\dagger}\rangle\rightarrow|a_{\mathrm{J}}^{\dagger}\rangle\rightarrow|a\rangle$,
where $|a_{\mathrm{J}}^{\dagger}\rangle(|b_{\mathrm{J}}^{\dagger}\rangle)$
holds the Stokes vector $\mathbf{S}^{a_{\mathrm{J}}^{\dagger}(b_{\mathrm{J}}^{\dagger})}=\mathbf{S}^{a(b)}-2(\mathbf{S}^{a(b)}\cdot\mathbf{S}^{\mathrm{J}})\mathbf{S}^{\mathrm{J}}$
\label{fig:S01JM}\bigskip{}
}
\end{figure}

\noindent From equation \eqref{A01}, it can found that clockwise
and anticlockwise walks on the sphere surface will induce opposite
values of solid angles. As shown in Fig.\textbf{~}\ref{Fig2.Solidangle}(b)
in main text, the spherical lune is shaped by two geodesics connecting
the antipodal states $|q_{1}\rangle$ and $|q_{2}\rangle$ passing
through $|a\rangle$ and $|b\rangle$ and forming a dihedral angle
$\psi_{\mathrm{B}}=\mathrm{arg}(\mu_{2})-\mathrm{arg}(\mu_{1})=\mathrm{arg}(\mu_{1}^{\ast}\mu_{2})$,
which is introduced by birefringent of optical element (also see Fig.~\ref{fig:S01JM}).
The corresponding lune area equals $2\psi_{\mathrm{B}}=\Omega_{abq_{1}}-\Omega_{abq_{2}}$.
In particular, when $|q_{1(2)}\rangle$ coincides with $|\mathrm{e}_{\pm}\rangle$,
$2\psi_{\mathrm{B}}$ also equals $4(\psi_{\mathrm{S}}^{b}-\psi_{\mathrm{S}}^{a})$.
It is easy to find that there is a relation $\Omega_{abb^{\dagger}a^{\dagger}}^{\mathrm{C}}=\Omega_{abq_{1}}+\Omega_{abq_{2}}$
and $\Omega_{abb^{\dagger}a^{\dagger}}^{\mathrm{J}}$ can be solved
by similar approach.

\section*{Appendix C: P2P Transformation on scalar vortex}

\subsection*{1. General P2P transformation}

The input scalar vortex ($|a^{+}\rangle$) is set as polarization
azimuth of $\psi_{\mathrm{S}}$, ellipticity of $\chi_{\mathrm{S}}$
and carrying OAM of $l$ 

\begin{equation}
|a^{+}\rangle=e^{il\phi}\mathbf{R}(\psi_{\mathrm{S}})\left(\begin{array}{c}
\cos(\chi_{\mathrm{S}})\\
i\sin(\chi_{\mathrm{S}})
\end{array}\right),\label{S22}
\end{equation}
where $\mathbf{R(\cdot)}$ is the standard rotation matrix. Then using
an optical element with Jones matrix $\mathbf{J}$ transfers $|a^{+}\rangle$
to the output vortex ($|b^{+}\rangle$) with a flipped handedness
and OAM charge $m$ as 

\begin{equation}
|b^{+}\rangle=e^{im\phi}\mathbf{R}(\psi_{\mathrm{S}})\left(\begin{array}{c}
\cos(\chi_{\mathrm{S}})\\
-i\sin(\chi_{\mathrm{S}})
\end{array}\right),\label{S23}
\end{equation}
then we can obtain the relation

\begin{equation}
e^{im\phi}\mathbf{R}(\psi_{\mathrm{S}})\left(\begin{array}{c}
\cos(\chi_{\mathrm{S}})\\
-i\sin(\chi_{\mathrm{S}})
\end{array}\right)=e^{il\phi}\mathbf{J}\mathbf{R}(\psi_{\mathrm{S}})\left(\begin{array}{c}
\cos(\chi_{\mathrm{S}})\\
i\sin(\chi_{\mathrm{S}})
\end{array}\right).\label{S24}
\end{equation}

For a scalar vortex, due to the rotation symmetry of the coordinate
choice for polarization azimuth, $\psi_{\mathrm{S}}=0$ is settled
for simplicity but without loss of generality, thus equation~\eqref{S24}
can be rewritten as 

\begin{equation}
e^{i\Delta l_{1}\phi}\left(\begin{array}{c}
\cos(\chi_{\mathrm{S}})\\
-i\sin(\chi_{\mathrm{S}})
\end{array}\right)=\mathbf{J}\left(\begin{array}{c}
\cos(\chi_{\mathrm{S}})\\
i\sin(\chi_{\mathrm{S}})
\end{array}\right),\label{S25}
\end{equation}
where $\Delta l_{1}=m-l$ is the variation of OAM from $|a^{+}\rangle$
to $|b^{+}\rangle$.

Generally, without considering chiral or magneto-optic materials,
$\mathbf{J}$ is linearly birefringent and the eigenstates will correspond
to linearly polarized orthogonal eigen-polarizations. Being unitary,
eigenvalues of $\mathbf{J}$ are given by complex exponentials of
$e^{i\psi_{\mathrm{D}}}\{e^{-i\psi_{\mathrm{B}}/2},e^{i\psi_{\mathrm{B}}/2}\}$
for dynamic phase delay $\psi_{\mathrm{D}}$ and birefringent phase
difference $\psi_{\mathrm{B}}$. And such two orthogonal eigen-polarizations
can be written as $\mathbf{R}(\psi_{\mathrm{R}})(1,0)^{\mathrm{\mathrm{T}}}$
and $\mathbf{R}(\psi_{\mathrm{R}})(0,1)^{\mathrm{T}}$, where $\psi_{\mathrm{R}}$
is the orientation angle of eigen-polarizations. Then, for such kinds
of optical elements, the Jones matrix $\mathbf{J}$ can be expressed
as

\begin{widetext}

\begin{equation}
\mathbf{J}=\left(\begin{array}{cc}
J_{1} & J_{2}\\
J_{3} & J_{4}
\end{array}\right)=e^{i\psi_{\mathrm{D}}}\left(\begin{array}{cc}
\cos\left(\frac{\psi_{\mathrm{B}}}{2}\right)-i\sin\left(\frac{\psi_{\mathrm{B}}}{2}\right)\cos\left(2\psi_{\mathrm{R}}\right) & -i\sin\left(\frac{\psi_{\mathrm{B}}}{2}\right)\sin\left(2\psi_{\mathrm{R}}\right)\\
-i\sin\left(\frac{\psi_{\mathrm{B}}}{2}\right)\sin\left(2\psi_{\mathrm{R}}\right) & \cos\left(\frac{\psi_{\mathrm{B}}}{2}\right)+i\sin\left(\frac{\psi_{\mathrm{B}}}{2}\right)\cos\left(2\psi_{\mathrm{R}}\right)
\end{array}\right).\label{S26}
\end{equation}

\end{widetext}

From equation~\eqref{S26}, it can be found that there are there
parameters $\{\psi_{\mathrm{D}}(\phi),\psi_{\mathrm{B}}(\phi),\psi_{\mathrm{R}}(\phi)\}$
to determine the Jones matrix for each $\phi$, while there are just
two equations to define their relations by equation~\eqref{S25}.
The result is that we have infinite choices to construct $\mathbf{J}$
to achieve the same transformation. That is, once a contribution from
dynamic phase delay $\psi_{\mathrm{D}}(\phi)$ is set, we always can
find a selection of $\{\psi_{\mathrm{B}}(\phi),\psi_{\mathrm{R}}(\phi)\}$.
Combining equations \eqref{S25} and \eqref{S26}, we can rewrite
$\mathbf{J}$

\begin{widetext}

\begin{equation}
\mathbf{J}=e^{i\psi_{\mathrm{D}}}\left(\begin{array}{cc}
\cos(\Delta l_{1}\phi-\psi_{\mathrm{D}})\cos(2\chi_{\mathrm{S}})+i\sin(\Delta l_{1}\phi-\psi_{\mathrm{D}}) & -i\cos(\Delta l_{1}\phi-\psi_{\mathrm{D}})\sin(2\chi_{\mathrm{S}})\\
-i\cos(\Delta l_{1}\phi-\psi_{\mathrm{D}})\sin(2\chi_{\mathrm{S}}) & \cos(\Delta l_{1}\phi-\psi_{\mathrm{D}})\cos(2\chi_{\mathrm{S}})-i\sin(\Delta l_{1}\phi-\psi_{\mathrm{D}})
\end{array}\right).\label{S26A}
\end{equation}

\end{widetext}

\noindent With equations \eqref{S26} and \eqref{S26A}, for any input
light with SOP of $\chi_{\mathrm{S}}$ to obtain a desired OAM variation
$\Delta l_{1}$, the design of three parameters $\{\psi_{\mathrm{D}}(\phi),\psi_{\mathrm{B}}(\phi),\psi_{\mathrm{R}}(\phi)\}$
can be solved for $\mathbf{J}$ of optical element.

To present the effect from $\{\psi_{\mathrm{D}},\psi_{\mathrm{B}}\}$
of the optical element, we give some comments and discussions on the
considered transformation. As shown in Fig.~\ref{fig:S01JM}, for
the input state $\left|a\right\rangle $, we can use a geodesic arc
join $\left|q_{1}\right\rangle $, $\left|a\right\rangle $ and $\left|q_{2}\right\rangle $
and let the arc go a rotation of $\psi_{\mathrm{B}}$ around the axis
defined by its eigen-polarizations $|q_{1(2)}\rangle$. Then the final
state $\left|b\right\rangle $ can be obtained on the corresponding
location as shown in Fig.~\ref{fig:S01JM}. From such a transformation,
the introduced dynamic phase is always equal to $\psi_{\mathrm{D}}$,
while the introduced geometric phase is $\Omega_{\mathrm{G}}/4=\Omega_{abb^{\dagger}a^{\dagger}}^{\mathrm{J}}\big/4$.
It is easy to find that the geometric phase equals $-\psi_{\mathrm{B}}/2$
if $\left|a\right\rangle $=$\left|q_{1}\right\rangle $ and $\psi_{\mathrm{B}}/2$
if $\left|a\right\rangle $=$\left|q_{2}\right\rangle $. Moreover,
if the input SOP is orthogonal to $\left|a\right\rangle $, i.e. the
antipodal point on the Poincar\'{e} sphere, the evolution will encircle
on an opposite direction but the same area using the same optical
element, this means that the introduced geometric phase is $-\Omega_{\mathrm{G}}/4$
for input light with orthogonal SOP.

\begin{figure*}
\includegraphics{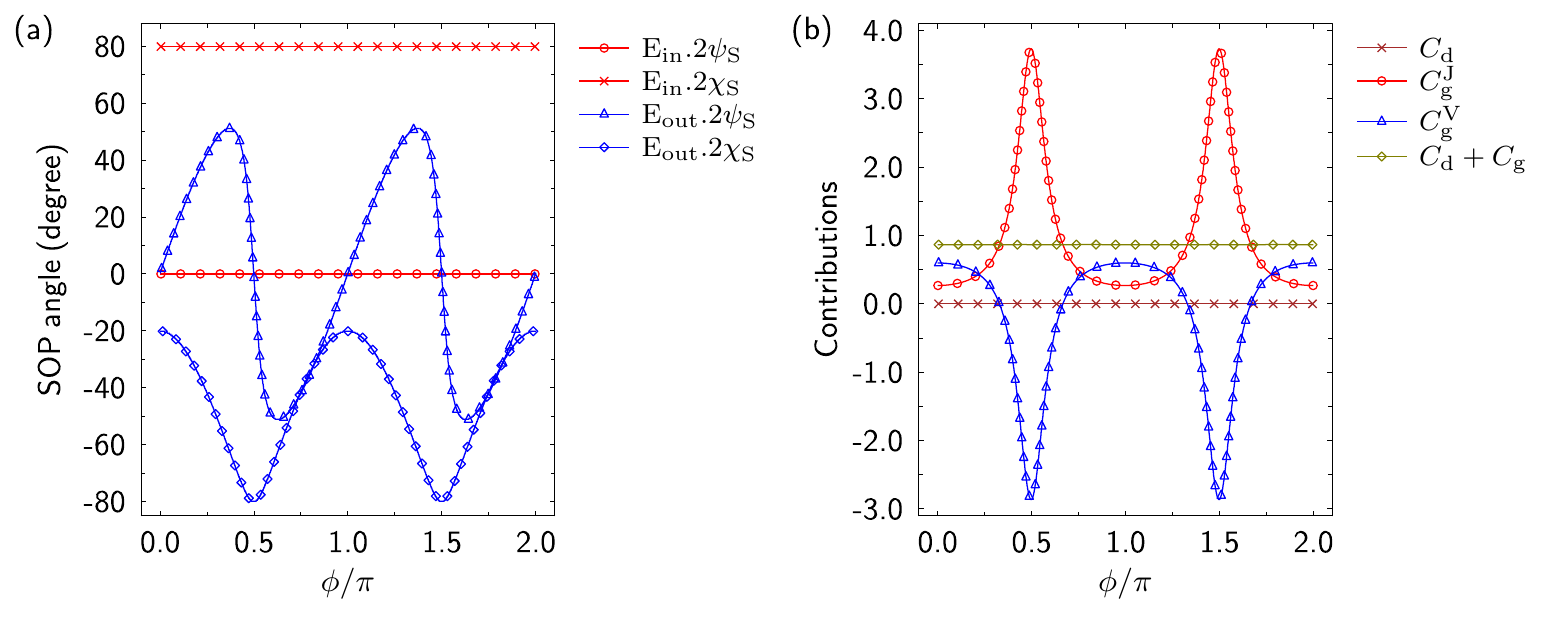}

\caption{Transformation with P2 for input light with SOP of $\{2\psi_{\mathrm{S}},2\chi_{\mathrm{S}}\}=\{0,80^{\circ}\}$.
(a) SOPs for input and output fields. (b) Detailed contributions for
OAM variation from each term. \label{fig:S02}\bigskip{}
}
\end{figure*}

\begin{figure}
\includegraphics{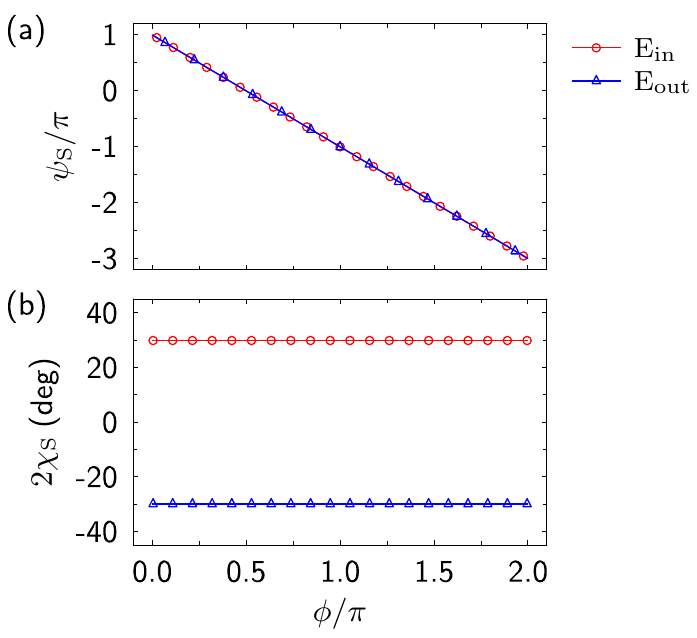}\caption{SOPs for input and output vector vortices shown in Fig.~\ref{Fig4.VVTrans}
in main text. (a) Azimuth angle $\psi_{\mathrm{S}}$ of the vector
vortices. (b) Ellipticity angle $\chi_{\mathrm{S}}$ of the vector
vortices. \label{fig:S03}}
\end{figure}

\subsection*{2. J-plates and q-plates}

For the J-plates \cite{Devlin2017S}, there is a special constrain
on P2P transformation. That is, the input field with an orthogonal
SOP and OAM charge of $k$ gives

\begin{equation}
|a^{-}\rangle=e^{ik\phi}\mathbf{R}(\psi_{\mathrm{S}})\left(\begin{array}{c}
i\sin(\chi_{\mathrm{S}})\\
\cos(\chi_{\mathrm{S}})
\end{array}\right),\label{S27}
\end{equation}
and the transferred output field with a flipped handedness and OAM
charge of $n$, yielding

\begin{equation}
|b^{-}\rangle=e^{in\phi}\mathbf{R}(\psi_{\mathrm{S}})\left(\begin{array}{c}
-i\sin(\chi_{\mathrm{S}})\\
\cos(\chi_{\mathrm{S}})
\end{array}\right).\label{S28}
\end{equation}

With the similar approach and setting $\psi_{\mathrm{S}}=0$, we can
find another relation for $\mathbf{J}$ as

\begin{equation}
e^{i\Delta l_{2}\phi}\left(\begin{array}{c}
-i\sin(\chi_{\mathrm{S}})\\
\cos(\chi_{\mathrm{S}})
\end{array}\right)=\mathbf{J}\left(\begin{array}{c}
i\sin(\chi_{\mathrm{S}})\\
\cos(\chi_{\mathrm{S}})
\end{array}\right),\label{S29}
\end{equation}
where $\Delta l_{2}=n-k$ is the variation of OAM from $|a^{-}\rangle$
to $|b^{-}\rangle$. Then, combining equations \eqref{S25} and \eqref{S29},
we can get the design parameters of $\mathbf{J}$. For a special and
simple case of input field with circular polarization ($2\chi_{\mathrm{S}}=\pi/2$),
the combination of equations \eqref{S25} and \eqref{S29} can reduce
$\mathbf{J}$ to

\begin{equation}
\left\{ \begin{aligned}J_{1} & =-J_{4}=\frac{1}{2}\left(e^{i\Delta l_{1}\phi}-e^{i\Delta l_{2}\phi}\right)\\
J_{2} & =J_{3}=-\frac{i}{2}\left(e^{i\Delta l_{1}\phi}+e^{i\Delta l_{2}\phi}\right)
\end{aligned}
\right..\label{S30}
\end{equation}

From equation~\eqref{S30}, we can find the eigenvalues as

\begin{equation}
\left\{ \begin{aligned}\mu_{1} & =i\exp\left(\frac{i(\Delta l_{1}+\Delta l_{2})\phi}{2}\right)\\
\mu_{2} & =-i\exp\left(\frac{i(\Delta l_{1}+\Delta l_{2})\phi}{2}\right)
\end{aligned}
\right.,\label{S31}
\end{equation}
and eigen-polarizations as

\begin{equation}
\left\{ \begin{aligned}|q_{1}\rangle & =\mathbf{R}\left(\frac{(\Delta l_{1}-\Delta l_{2})\phi-\pi}{4}\right)\left(\begin{array}{c}
1\\
0
\end{array}\right)\\
|q_{2}\rangle & =\mathbf{R}\left(\frac{(\Delta l_{1}-\Delta l_{2})\phi-\pi}{4}\right)\left(\begin{array}{c}
0\\
1
\end{array}\right)
\end{aligned}
\right.,\label{S32}
\end{equation}
thus the parameters $\{\psi_{\mathrm{D}}(\phi),\psi_{\mathrm{B}}(\phi),\psi_{\mathrm{R}}(\phi)\}$
for $\mathbf{J}$ can be found as

\begin{equation}
\left\{ \begin{aligned}\psi_{\mathrm{D}}(\phi) & =\frac{(\Delta l_{1}+\Delta l_{2})\phi}{2}\\
\psi_{\mathrm{B}}(\phi) & =\pi\\
\psi_{\mathrm{R}}(\phi) & =\frac{(\Delta l_{1}-\Delta l_{2})\phi-\pi}{4}
\end{aligned}
\right..\label{S33}
\end{equation}
It can be found that this kind of J-plates is a half-plate but possesses
space-variant dynamic phase delay and orientation angle and can transform
scalar vortices with circular polarizations. Further, q-plates would
be as a special case of J-plate with opposite variation of OAM, i.e.
$\Delta l_{1}=-\Delta l_{2}$. Thus, for the q-plate, there is only
the contribution from geometric phases while none from dynamic phases.
It should be noted that, for J-plates and q-plates, the geometric
contributions actually come from the third term of equation \eqref{S21}.
And to calculate the OAM variation, $\psi_{\mathrm{S}}$ should be
calculated using equation \eqref{S17} while not equation \eqref{S4a}
due to there is a singularity for circularly polarized fields. And
this singularity also induces the geometric contribution coming from
$C_{\mathrm{g}}^{\mathrm{V}}$ while not $C_{\mathrm{g}}^{\mathrm{J}}$,
which is different from the demonstration of P2P transformation in
main text. However, this phenomenon occurs just because the circularly
polarized fields are selected as the reference fields and it can be
resolved if another pair of reference fields are adopted.

\subsection*{3. Spiral phase plates (SPPs)}

\paragraph{Type I: }

This type of spiral phase plates (SPP-I) is fabricated with homogeneous
materials and can provide required phase delay by designing path length.
In the transformation, only dynamic phases $\psi_{\mathrm{D}}(\phi)$
should be considered since $\psi_{\mathrm{B}}(\phi)=0$. From equation
\eqref{S26}, the corresponding Jones matrix for this kind of spiral
phase plates can be written

\begin{equation}
\mathbf{J}=e^{i\psi_{\mathrm{D}}}\left(\begin{array}{cc}
1 & 0\\
0 & 1
\end{array}\right).
\end{equation}

So they can not change the SOP but always introduce the same dynamic
phase for any SOP. That is to say the same OAM variation can be achieved
for any input fields.

\paragraph{Type II: }

There is another kind of spiral phase plates (SPP-II), which can provide
a desired OAM variation only for a specific SOP of input light beam
and the output possessing the same SOP. Actually, the input SOP is
just coincided with one of eigen-polarizations so that the dynamic
or geometric phase or both of them would contribute to the OAM variations.
For simplicity, we set the input SOP as $\{2\psi_{\mathrm{S}},2\chi_{\mathrm{S}}\}=\{0,0\}$
(i.e., $\psi_{\mathrm{R}}(\phi)=0$). Thus, with equation \eqref{S26},
we can obtain the Jones matrix

\begin{equation}
\mathbf{J}=e^{i\psi_{\mathrm{D}}}\left(\begin{array}{cc}
e^{-i\psi_{\mathrm{B}}/2} & 0\\
0 & e^{i\psi_{\mathrm{B}}/2}
\end{array}\right).
\end{equation}

It can found that all the contribution comes from geometric phase
if $\psi_{\mathrm{D}}(\phi)=0$. For this case, the orthogonal input
SOPs will get opposite OAM variations. There is another extreme case
of $\psi_{\mathrm{B}}(\phi)=0$, which is exactly the SPP-I plate.
Overall, this type of plates actually is a common P2P transformation
plate with the same input and output SOPs.

\section*{Appendix D: Vortex beams in numerical simulations}

\noindent For general vector beams, such as cylindrical vortices,
the field can be expressed as

\begin{widetext}

\begin{equation}
\mathbf{E}\left(\phi\right)=\frac{1}{\sqrt{2}}\cos\left(\frac{\pi}{4}-\chi_{\mathrm{S}}\right)(\hat{\mathbf{x}}+i\mathbf{\hat{y}})e^{i(l_{\mathrm{L}}\phi-\psi_{\mathrm{S}})}+\frac{1}{\sqrt{2}}\sin\left(\frac{\pi}{4}-\chi_{\mathrm{S}}\right)(\mathbf{\hat{x}}-i\mathbf{\hat{y}})e^{i(l_{\mathrm{R}}\phi+\psi_{\mathrm{S}})},\label{EqVVF}
\end{equation}

\end{widetext}

\noindent where $l_{\mathrm{L}}$ and $l_{\mathrm{R}}$ are topological
charges of field components with left- and right-handed circular polarization,
respectively. With equation \eqref{EqVVF}, there is $I_{\mathrm{E}}(\phi)=1$.
For a scalar vortex beam with topological charge $l$, it is easy
to be obtained by setting $l_{\mathrm{L}}=l_{\mathrm{R}}=l$. For
P2P transformation shown in Fig.~\ref{Fig3.P2PTrans}(c) in the main
text, we set the input light with SOP of $\{2\psi_{\mathrm{S}},2\chi_{\mathrm{S}}\}=\{0,50^{\circ}\}$
and topological charge of $l_{\mathrm{L}}=l_{\mathrm{R}}=0$ or $1$.
For the transformation shown in Fig.~\ref{Fig3.P2PTrans}(d) in the
main text, we set the input light with $\{2\psi_{\mathrm{S}},2\chi_{\mathrm{S}}\}=\{0,80^{\circ}\}$
and $l_{\mathrm{L}}=l_{\mathrm{R}}=0$, where the transformation with
P2 is detailed in Fig.~\ref{fig:S02}. And for the transformation
on vector vortex shown in Fig.~\ref{Fig4.VVTrans} in the main text,
we set the input light with SOP of $\{2\psi_{\mathrm{S}},2\chi_{\mathrm{S}}\}=\{2\pi-4\phi,\pi/6\}$
and topological charge of $\{l_{\mathrm{L}},l_{\mathrm{R}}\}=\{1,-3\}$,
and the detailed SOPs of input and output vector vortices are presented
in Fig.~\ref{fig:S03}.

\bigskip{}


%

\noindent 
\end{document}